\documentclass[letterpaper]{article} 
\usepackage{aaai2026}  
\usepackage{times}  
\usepackage{helvet}  
\usepackage{courier}  
\usepackage[hyphens]{url}  
\usepackage{graphicx} 
\usepackage{natbib}  
\usepackage{caption} 
\usepackage{tabularx}
\usepackage{amssymb}
\usepackage{pifont}
\newcommand{\xmark}{\ding{55}}%
\usepackage{algorithm}
\usepackage{algorithmic}
\usepackage{booktabs}
\usepackage{makecell}
\usepackage{threeparttable}
\usepackage{newfloat}
\usepackage{listings}
\DeclareCaptionStyle{ruled}{labelfont=normalfont,labelsep=colon,strut=off} 
\floatstyle{ruled}
\newfloat{listing}{tb}{lst}{}
\floatname{listing}{Listing}
\title{NewsTorch: A PyTorch-based Toolkit for Learner-oriented News Recommendation}
\author {
    Rongyao Wang\textsuperscript{\rm 1},
    Veronica Liesaputra\textsuperscript{\rm 1},
    Zhiyi Huang\textsuperscript{\rm 1}
}
\affiliations {
    \textsuperscript{\rm 1}School of Computing, University of Otago, Dunedin 9016, New Zealand\\
    rongyao.wang@postgrad.otago.ac.nz, veronica.liesaputra@otago.ac.nz, zhiyi.huang@otago.ac.nz
}
\usepackage{bibentry}

\begin{document}

\maketitle

\begin{abstract}
News recommender systems are devised to alleviate the information overload, attracting more and more researchers' attention in recent years.
The lack of a dedicated learner-oriented news recommendation toolkit hinders the advancement of research in news recommendation.
We propose a PyTorch-based news recommendation toolkit called NewsTorch, developed to support learners in acquiring both conceptual understanding and practical experience. 
This toolkit provides a modular, decoupled, and extensible framework with a learner-friendly GUI platform that supports dataset downloading and preprocessing.
It also enables training, validation, and testing of state-of-the-art neural news recommendation models with standardized evaluation metrics, ensuring fair comparison and reproducible experiments. 
Our open-source toolkit is released on Github: https://github.com/whonor/NewsTorch.  
\end{abstract}


\section{Introduction}
\textbf{Background}: 
News recommender system is one type of recommender systems, which is able to recommend interested news content to users and reduce the impact of the information overload problem.
Various news online platforms have implemented news recommender systems, such as Bing News, CNN and BBC.
News recommendation has become a vital role in influencing users' reading behaviours in the era of artificial intelligence \cite{wang2024trustworthy}.

\textbf{Motivation}: 
In recent years, deep learning technology has achieved promising advancements, including generative large language models (LLMs) and artificial intelligence agents.
Some superior deep learning-based methods \cite{wu2019nrms, wang2022news, wang2023intention, wu2021empowering} are utilized in news recommender systems. 
News recommendation has attracted increasing attention from researchers, resulting in a large number of publications in recent years \cite{wu2024survey}. 
However, studying news recommendation remains challenging for beginners due to complex data preparation, difficulties in reproducing code, and heterogeneous evaluation frameworks.

\textbf{Challenge}: 
The news recommendation toolkit is a valuable resource that facilitates experimental research by providing a fair evaluation platform across different baselines.
Most open-source news recommendation libraries \cite{iana2023newsreclib} devise a unified and highly configurable framework for researchers built on a high-level library such as PyTorch-Lightning\footnote{https://lightning.ai/pytorch-lightning}, Hydra\cite{Yadan2019Hydra}.
While existing high-level libraries support reproducible research and comprehensive experiments, they provide limited support for learners. 
This motivates the design of a learner-friendly toolkit for news recommendation studies.
Therefore, we propose NewsTorch, a framework designed to facilitate model reproduction, dataset preprocessing, and experimental operations with ease for learners.
The essential differences between NewsTorch and other popular news recommendation libraries are demonstrated in Table \ref{tab:comparison}.

\textbf{Contribution}: 
(1) In this paper, we propose a learner-oriented news recommendation toolkit named NewsTorch, which includes a unified framework and a learner-friendly GUI (Graph User Interface) to configure, train, and evaluate all baselines.
(2) We reproduce a wide range of neural news recommendation models, including deep learning-based models, GNN-based models, and LLM-based news recommendation models. Moreover, NewsTorch allows learners to flexibly customize models and datasets via an independent configuration mechanism.
(3) To the best of our knowledge, NewsTorch is the first learner-oriented news recommendation toolkit to support both GNN-based and LLM-based models within a decoupled framework.

\begin{figure*}[htbp]  
  \centering
  \includegraphics[width=0.9\textwidth]{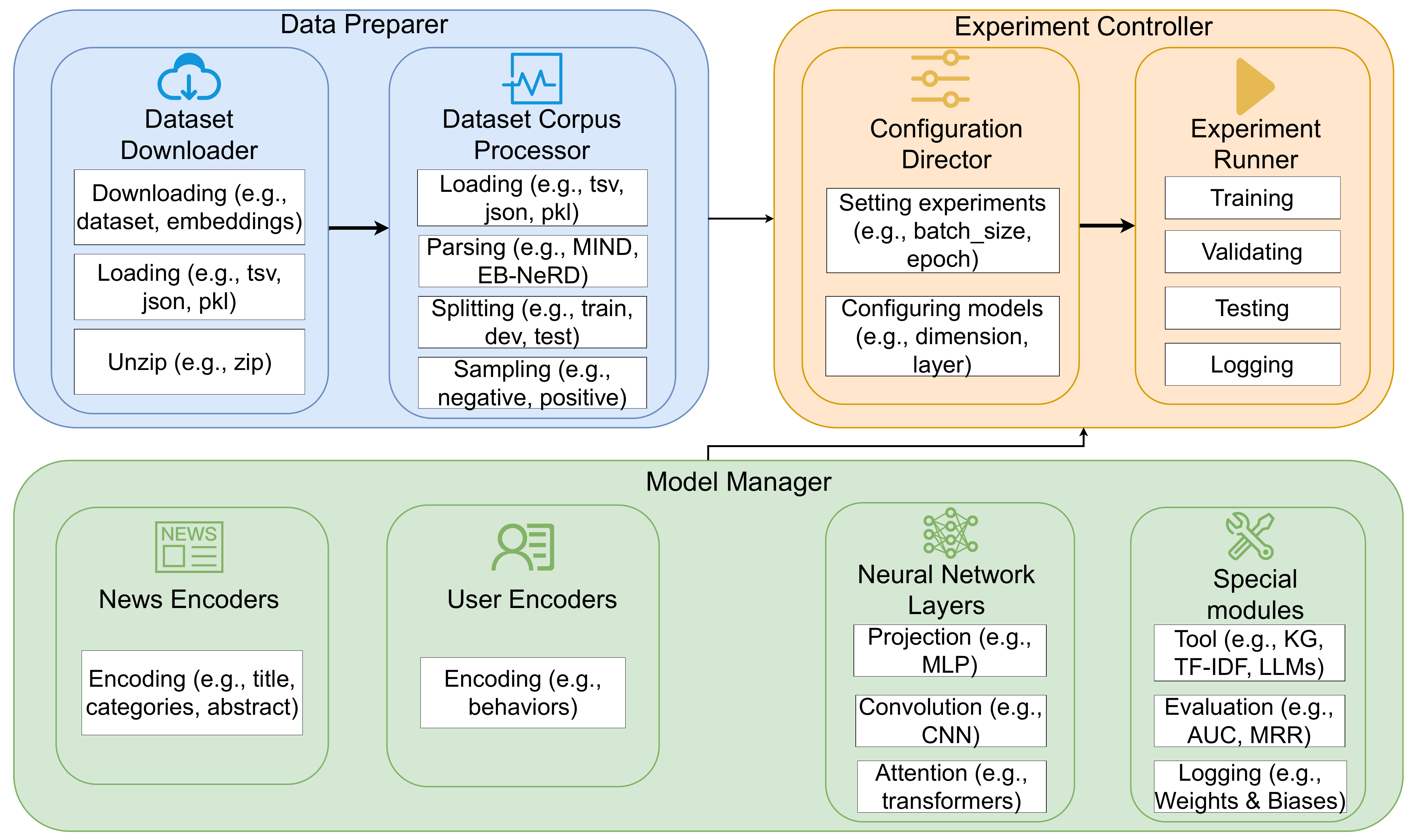}
  \caption{The framework of the NewsTorch toolkit.}
  \label{fig:framework}
\end{figure*}

\section{Overview of NewsTorch}
In this section, we introduce the framework and components of NewsTorch.
The framework is illustrated in Figure \ref{fig:framework}, which is composed of different functional components: \textbf{Dataset Preparer}, \textbf{Experiment Controller}, \textbf{Model Manager} and \textbf{Web GUI}.

\textbf{Dataset Preparer}. This component is devised to download and process different news recommendation datasets, including \textbf{Dataset Downloader} and \textbf{Dataset Corpus Processor}. 
The \textbf{Dataset Downloader} is responsible for acquiring datasets, whereas \textbf{Dataset Corpus Processor} enables the transformation of diverse news recommendation datasets into a unified format.
Various popular news recommendation datasets have been employed in our toolkit, such as MIND\cite{wu2020mind} and EB-NeRD\cite{kruse2024eb}, which can be easily utilized by learners in their experiments.

\textbf{Experiment Controller}. This component enables users to independently configure models and experiments and to initiate experimental runs, comprising \textbf{Configuration Director} and \textbf{Experiment Runner}.
Specifically, \textbf{Configuration Director} is utilized to assist users in customizing parameters with an independent configuration mechanism, using YAML files to configure each model within a dedicated directory.
\textbf{Experiment Runner} is designed to execute different neural news recommendation models, supporting independent training, validation, and testing.
In order to record experiments' status and results, we introduce Weights\&Biases \footnote{https://wandb.ai/site} into \textbf{Experiment Runner}.

\textbf{Model Manager}. We design this component to manage all models' codes, including news encoders, user encoders, various neural network layers, and special modules.
Sufficient and diverse news recommendation models are covered in our toolkit, such as deep learning-based models, GNN-based models, and LLM-based models.
Moreover, adding new models to this framework is straightforward due to its modular and decoupled design.

\textbf{Web GUI}. This provides various functions that assist learners in directly downloading and processing datasets. 
After preprocessing, they can click the 'start' button to train, validate, and test models. 


\begin{table}[t]
\setlength{\abovecaptionskip}{2pt}
\setlength{\belowcaptionskip}{0pt}
\setlength{\intextsep}{5pt}
\centering
\scriptsize
\renewcommand{\arraystretch}{1.1}
\begin{threeparttable}
\begin{tabularx}{\columnwidth}{lXXX}
\toprule
\textbf{Feature} & \textbf{News-Recommendation\tnote{a}} & \textbf{NewsRecLib\tnote{b}} & \textbf{NewsTorch} \\
\midrule
Models & DL\tnote{c} & DL & DL,GNN,LLM \\
\midrule
Multi-Dataset & \xmark & \checkmark & \checkmark \\
\midrule
Active & \xmark & \checkmark & \checkmark \\
\midrule
Multi-GPU & \xmark & \xmark & \checkmark \\
\midrule
GUI & \xmark & \xmark & \checkmark \\
\bottomrule
\end{tabularx}
\begin{tablenotes}
\footnotesize
\item[a] \url{https://github.com/yusanshi/news-recommendation}
\item[b] \url{https://github.com/andreeaiana/newsreclib}
\item[c] {Deep learning-based methods}
\end{tablenotes}
\end{threeparttable}
\caption{Compact comparison of news recommendation libraries.}
\label{tab:comparison}
\end{table}

\section{Conclusion}
NewsTorch is a neural news recommendation toolkit to support learners in acquiring both conceptual understanding and practical experience.
It encourages a fair experimental environment and a unified framework with a learner-friendly GUI.
With a modular and extensible architecture, the toolkit allows learners to flexibly customize models and datasets via an independent configuration mechanism.
Moreover, NewsTorch supports advanced functionalities, including multi-GPU training, experiment tracking, the implementation of diverse models, and the processing of multiple datasets.
In the future, we plan to explore more advanced functions and continue updating our toolkit.
New state-of-the-art news recommendation models and datasets will be incorporated into NewsTorch.
 

\section{Acknowledgments}

\bigskip
\noindent This work is supported by the University of Otago - China Scholarship Council Doctoral Scholarship, grant number 202308370189. 
We would also like to thank the anonymous AAAI reviewers for their positive feedback and helpful suggestions.

\bibliography{aaai2026}


\end{document}